\documentclass{article}
\usepackage{dcase2024,amsmath,graphicx,xurl,times,booktabs, tabularx,comment}

\usepackage{hhline}
\usepackage{hyperref}
\usepackage{xcolor}
\usepackage{cite}
\usepackage{multirow,url,hyperref}
\usepackage{siunitx}

\usepackage{pifont}
\usepackage{multirow}

\usepackage{hyperref}
\usepackage{xcolor}

\usepackage{etoolbox}
\makeatletter
\pretocmd{\thebibliography}{\footnotesize}{}{}
\makeatother

\title{Data-Efficient Low-Complexity Acoustic Scene Classification \\ in the DCASE 2024 Challenge}

\name{Florian Schmid$^{1}$,
       Paul Primus$^{1}$,
       Toni Heittola$^{3}$,
       }
 \secondlinename{	  
       Annamaria Mesaros$^{3}$,
       Irene Martín-Morató$^{3}$,
       Khaled Koutini$^{2}$, 
       Gerhard Widmer$^{1,2}$
       }
 \address{$^1$Institute of Computational Perception, Johannes Kepler University Linz, Austria \\
 $^2$LIT Artificial Intelligence Lab, Linz, Austria,  $^3$Computing Sciences, Tampere University, Finland\\
 \{florian.schmid, paul.primus, gerhard.widmer\}@jku.at\\
 \{toni.heittola, annamaria.mesaros, irene.martinmorato\}@tuni.fi
  }

\begin{document}

\ninept
\maketitle

\begin{sloppy}

\begin{abstract}
This article describes the \textit{Data-Efficient Low-Complexity Acoustic Scene Classification} Task in the DCASE 2024 Challenge and the corresponding baseline system. The task setup is a continuation of previous editions (2022 and 2023), which focused on recording device mismatches and low-complexity constraints. This year's edition introduces an additional real-world problem: participants must develop data-efficient systems for five scenarios, which progressively limit the available training data. The provided baseline system is based on an efficient, factorized CNN architecture constructed from inverted residual blocks and uses Freq-MixStyle to tackle the device mismatch problem. The task received 37 submissions from 17 teams, with the large majority of systems outperforming the baseline. The top-ranked system's accuracy ranges from 54.3\% on the smallest to 61.8\% on the largest subset, corresponding to relative improvements of approximately 23\% and 9\% over the baseline system on the evaluation set. 

\end{abstract}


\vspace{-4pt}
\section{Introduction}
\label{sec:intro}
\vspace{-4pt}

Acoustic Scene Classification (ASC) aims at detecting the environmental context in which audio was captured, based on a short excerpt~\cite{benetos2018approaches}. The environmental context is given as a set of pre-defined acoustic scene classes such as \textit{Metro station}, \textit{Urban park}, or \textit{Public square}. Since its inception, the ASC task has been an integral part of the DCASE Challenge. Each year's edition focused on one or multiple challenging machine-learning aspects in addition to the supervised classification task itself. These aspects include open-set classification~\cite{Mesaros2019asc_in_dcase19}, 
constraints on the model's size and computational complexity~\cite{heittola20asc_in_dcase20, Martin21asc_in_dcase21, Martin22asc_in_dcase22}, and generalization across different recording devices~\cite{Mesaros18multi_dev_dataset, heittola20asc_in_dcase20}. These additional problems target the real-world applicability of ASC systems; for instance, the methods should be robust to diverse recording devices and sufficiently lightweight to be deployable on embedded devices. In the 2024 edition\footnote{Task Description Page: \url{https://dcase.community/challenge2024/task-data-efficient-low-complexity-acoustic-scene-classification}} of the ASC task, an additional challenging real-world aspect is addressed: the limited availability of training data. This setting intends to spark research on data-efficient learning methods capable of achieving high classification performance given only a small number of labeled acoustic scene examples for training.

\begin{figure}[t!]
\centering
{\includegraphics[width=\linewidth]{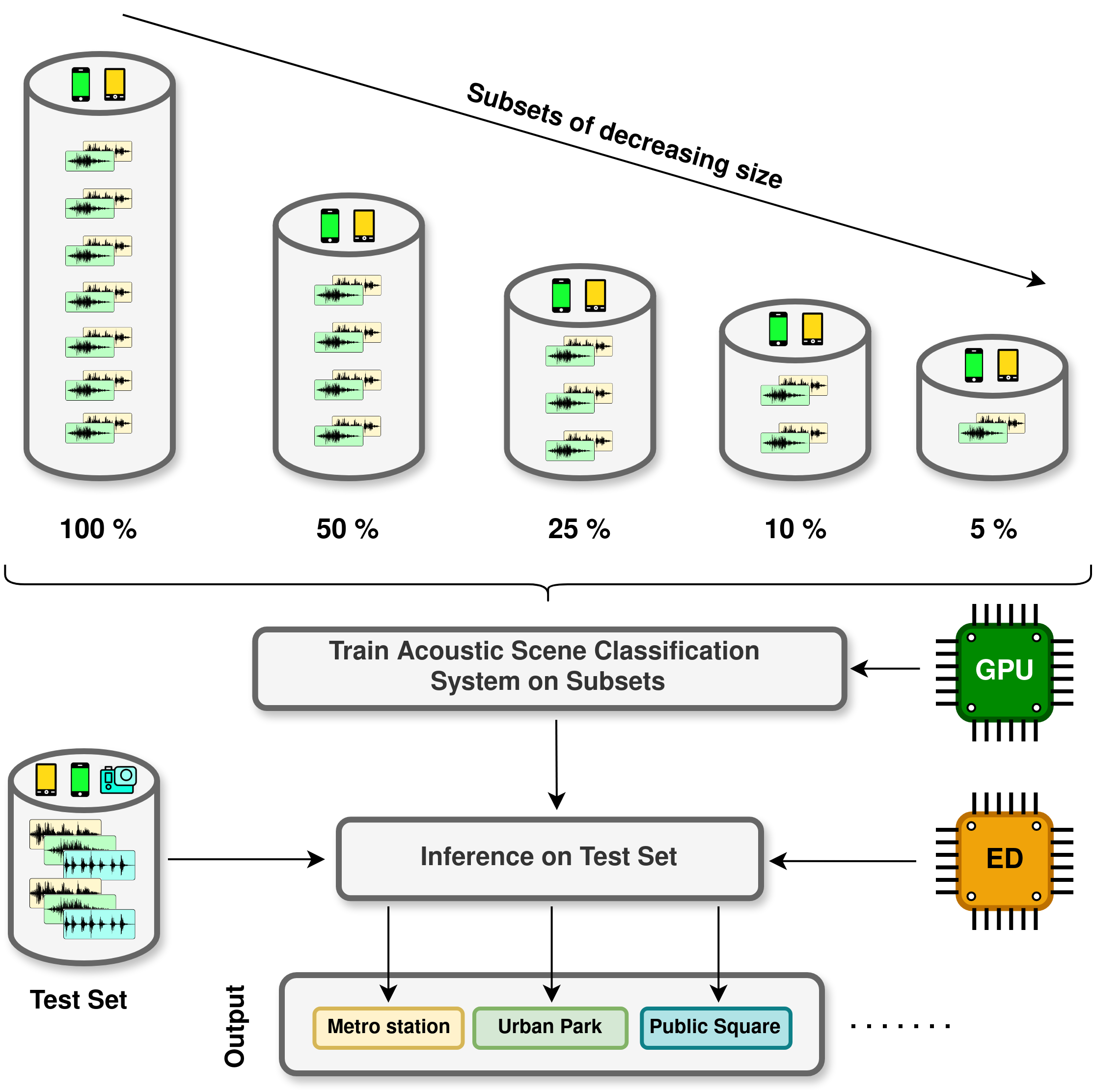}}
\caption{Overview of Data-Efficient Low-Complexity Acoustic Scene Classification. Submitted systems must be trained on five datasets of varying sizes, they must generalize to unseen recording devices, and they are required to be lightweight enough for inference on an embedded device (ED).}
\vspace{-12pt}
\label{fig:overview}
\end{figure}

Figure~\ref{fig:overview} shows an overview of the task setup. The ASC systems must be trained on subsets of a fixed training set that progressively limit the number of training samples, where the smallest subset only contains 5\% of the audio snippets in the full training set (see Section ~\ref{subsec:data-efficient}). The training procedure is not limited in terms of complexity and may be executed on high-end GPU hardware. However, aligned with real-world requirements, the system must be lightweight for inference such that it can be deployed on embedded devices (see Section ~\ref{subsec:complexity}). Additionally, the developed ASC system must be robust to unseen recording devices. To test this ability, the test set includes audio clips recorded by new devices that are not available in the training sets (see Section ~\ref{subsec:data-efficient}).

The rest of this paper is structured as follows: Section~\ref{sec:previous} briefly discusses the role of low-complexity constraints and the device generalization problem in previous editions of the task. Section~\ref{sec:setup} introduces the setup for Data-Efficient Low-Complexity Acoustic Scene Classification in the DCASE 2024 Challenge; the baseline system is presented in Section~\ref{sec:bl_system}. The outcome of the challenge is analyzed in Section~\ref{sec:results} and the paper is concludes in Section~\ref{sec:conclusion}.

\vspace{-4pt}
\section{Previous Editions}
\label{sec:previous}
\vspace{-4pt}

The low-complexity aspect has already been investigated in previous DCASE challenges and has undergone several refinements. In the 2020~\cite{heittola20asc_in_dcase20} and 2021~\cite{Martin21asc_in_dcase21} editions, systems were limited with respect to model size, allowing 500 kB and 128 kB for non-zero parameters, respectively. In the 2022 edition~\cite{Martin22asc_in_dcase22}, the complexity constraint additionally included computational complexity, allowing a maximum of 30 MMACs (million multiply-accumulate operations), modeled after Cortex-M4 devices. The maximum number of parameters was 128K, with the variable type fixed to INT8. The 2023 edition took this one step further and included model size and computational complexity as part of the ranking metric, requiring participants to tune the system's performance--complexity trade-off. In response to the low-complexity requirements, training techniques such as Sparsification~\cite{Yang2021lottery}, Pruning~\cite{Koutini2021CPJKU21}, Quantization~\cite{Kim2021qti}, or Knowledge Distillation~\cite{Schmid2022CPJKU22} have been extensively studied, and efficient factorized CNN architectures~\cite{Tan2023a, Cai2023a, Schmid2023CPJKU23} have been designed.

Besides low-complexity techniques, substantial research has been conducted on the device mismatch problem. Efforts to improve device generalization involved suppressing device information via normalization~\cite{Kim2021qti} and domain adaptation~\cite{Koutini2020CPJKU20}, balancing the devices by changing the sampling distribution~\cite{Lee2022} and augmenting audio segments with device translators~\cite{Kim2021qti}, Freq-MixStyle~\cite{Kim2022FMS, Schmid2022CPJKU22}, and device impulse response augmentation~\cite{Morocutti23DIR}.

\vspace{-4pt}
\section{Task Setup}
\label{sec:setup}
\vspace{-4pt}

While low complexity and generalization across different recording devices are well-studied topics, the specific aspect of interest in the 2024 edition is the limited availability of acoustic scene data for training. Specifically, participants were encouraged to develop data-efficient systems and study techniques that can alleviate the data scarcity problem, such as using extensive audio augmentation methods, transferring knowledge from general-purpose audio datasets, or incorporating well-suited inductive biases.

\vspace{-3pt}
\subsection{Dataset}
\vspace{-3pt}

The task builds on top of the \textit{TAU Urban Acoustic Scenes 2022 Mobile} dataset~\cite{Mesaros18multi_dev_dataset, heittola20asc_in_dcase20}, which was used in the 2022 and 2023 editions of the task~\cite{Martin22asc_in_dcase22}. The dataset provides one-second audio snippets with a sampling rate of 44.1 kHz in single-channel, 24-bit format and consists of recordings from ten distinct acoustic scenes.


The audio was recorded in multiple European cities with four recording devices in parallel. The primary device, referred to as device \textit{A}, is a high-quality binaural device, while \textit{B}, \textit{C} and \textit{D} are commonly available consumer devices. Additionally, 10 simulated devices (\textit{S1-S10}) are created using audio from device A and a set of impulse responses from mobile devices. For details on the dataset creation and the exact distribution of devices, please refer to \cite{heittola20asc_in_dcase20}.

The data is split into a development and an evaluation set. The development set, consisting of 64 hours of audio, contains 3 real devices (A, B, C) and 6 simulated devices (S1--S6). The evaluation set comprises five unseen devices (D and S7-S10) and two unseen cities, in addition to devices and cities overlapping with the development set. The evaluation set is used to rank submissions and therefore comes without corresponding scene labels. Device and city information is not provided for recordings in the evaluation set.

\begin{table*}[t!]
\begin{center}
\begin{tabular}{@{}l|cc|cc|llll@{}}
\toprule
\textbf{System Label} & \textbf{Score} & \textbf{Team Rank} & \textbf{Size} & \textbf{MACs} & \textbf{Architecture} & \textbf{Complexity} & \textbf{Dev. Gen.} & \textbf{External} \\ \midrule
 Han\_SJTUTHU\_task1\_2 & 58.2 & \textbf{1} & 126kB & 29M & SSCP-Mobile & fp16, KD, prun. & FMS & PaSST \\
 Shao\_NEPUMSE\_task1\_1 & 57.2 & \textbf{3} & 107kB & 16M & IRMamba & int8, KD & FMS, DIR & PaSST \\  
MALACH24\_JKU\_task1\_1 & 57.0 & \textbf{2} & 122kB & 29M & CP-Mobile & fp16, KD & FMS, DIR & AudioSet \\  
Yeo\_NTU\_task1\_2 & 56.1 & \textbf{5} & 122kB & 29M & CP-Mobile & fp16, KD & FMS, DIR & PaSST \\
 Cai\_XJTLU\_task1\_3 & 56.0 & \textbf{4} & 126kB & 29M & TF-SepNet & int8, KD & FMS, DIR & AudioSet \\
Park\_KT\_task1\_2 & 55.4 & \textbf{6} & 126kB & 26M & GhostRes2Net & fp16, KD & FMS, DIR & PaSST, EAT \\
OO\_NTUPRDCSG\_task1\_1 & 54.8 & \textbf{7} & 116kB & 29M & MofleNet & int8 & FMS, DIR & - \\
Werning\_UPBNT\_task1\_1 & 54.4 & \textbf{8} & 122kB & 29M & CP-Mobile & fp16, KD & FMS & AudioSet \\
Truchan\_LUH\_task1\_1 & 53.1 & \textbf{9} & 94kB & 29M & Isotropic & fp16 & FMS, DIR & - \\
Yan\_NPU\_task1\_1 & 52.9 & \textbf{10} & 124kB & 29M & MAR-CNN & fp32 & FMS & - \\ \midrule
Baseline & 50.7 & & 122kB & 29M & CP-Mobile & fp16 & FMS & - \\
\bottomrule
\end{tabular}
\caption{This table lists the top-ten teams' best systems according to their evaluation set performance. \textbf{Team Rank} indicates the team's overall rank, which is based on multiple submitted systems, and \textbf{Score} is the average accuracy across all splits of the respective system listed in the table. int8, fp16, and fp32 refer to the numerical precision of model parameters for inference, corresponding to 8, 16, and 32 bits, respectively. KD, FMS, and DIR are abbreviations for Knowledge Distillation, Freq.-MixStyle, and Device Impulse Response augmentation, respectively, and the column \textbf{External} indicates external resources used.} 
\label{tab:results}
\end{center}
\vspace{-16pt}
\end{table*}

\vspace{-3pt}
\subsection{Data-Efficient Evaluation}
\label{subsec:data-efficient}
\vspace{-3pt}

The development set used for the 2024 challenge is the same one as used in the previous two years and described above. It comes with the same pre-defined split into a development-train and a development-test partition. The development-train set contains six devices (A, B, C, S1-S3), leaving three unseen devices (S4-S6) for the development-test set to measure the device generalization performance.  

For the evaluation of data efficiency, this year's setup introduces five pre-defined subsets that progressively limit the available training data and contain 100\%, 50\%, 25\%, 10\%, and 5\% of the recordings in the development-train set. The distribution of acoustic scenes, cities, and recording devices is kept similar across all subsets. The smaller subsets are fully included in the larger ones, corresponding to the idea of progressively collecting more data. 

Participants are allowed to submit up to three different systems that may be specialized for the different training set sizes. Each system must be trained on all five subsets, and the performances on the development-test set must be reported. A system is considered to be the same if its architecture and design choices (such as building blocks, features, data augmentation techniques, decision-making, etc.) remain the same. However, basic hyperparameters like the number of update steps, learning rate, batch size, or regularization strength may vary for training on the different subsets.

All systems must be trained only on the respective subset and the explicitly allowed external resources. The allowed external resources include general-purpose audio datasets, such as AudioSet~\cite{Gemmeke17audioset} or FSD50K~\cite{Fonseca22fsd50k}, but no datasets specific to acoustic scenes. 

The leaderboard ranking score is computed as follows. First, class-wise macro-averaged accuracies for all $P=5$ development-train subsets and all $N$ submissions are computed. The accuracy of the $n$-th submission on the $p$\% subset is denoted as $ACC_{n,p}$. The scores are then aggregated by choosing the best-performing system for each subset and averaging the resulting accuracies. 

\vspace{-3pt}
\begin{equation}
    \mathrm{score} := \frac{1}{P} \sum_{p \in \{5,10,25,50,100\}} \max_{n \in \{ 1,...,N\}} ACC_{n,p}
\end{equation}
\vspace{-3pt}

The outlined setup encourages research into the following scientific questions: how does the performance of systems vary with the number of available labeled training samples? how can systems be adapted to better cope with the limited availability of labeled training data? how can general-purpose audio datasets be exploited to mitigate the need for larger amounts of acoustic scenes?

\vspace{-3pt}
\subsection{System Complexity Requirements}
\label{subsec:complexity}
\vspace{-3pt}

The system complexity is limited in terms of model size and MMACs. The maximum memory allowance for model parameters is 128 kB, with no requirement regarding the numerical representation. That is, participants can trade off the number of parameters and the numerical representation. For example, the memory limit translates to 128K parameters when using 8-bit quantization, or 32K parameters when using 32-bit precision. The computational complexity is limited to 30 MMACs for the inference on a one-second audio segment. These complexity limits are modeled after Cortex-M4 devices (e.g., STM32L496@80MHz or Arduino Nano 33@64MHz).


\vspace{-4pt}
\section{Baseline System}
\label{sec:bl_system}
\vspace{-4pt}

The baseline system is a simplified version of the top-ranked system submitted to the 2023 edition~\cite{Schmid2023cpm}. It is based on a receptive-field-regularized, factorized CNN design. Audio input is resampled to 32 kHz and converted to mel spectrograms using a 4096-point FFT with a window size of 96 ms and a hop size of approximately 16 ms, followed by a mel transformation with a filterbank of 256 mel bins. The system is trained for 150 epochs using the AdamW optimizer and a batch size of 256. Freq-MixStyle~\cite{Kim2022FMS,Schmid2022CPJKU22} is applied to tackle the device mismatch problem, and time rolling of the waveform and frequency masking are used to augment the training data. The baseline system requires 29.4 MMACs for the inference on a one-second audio clip. The memory required for the model parameters amounts to 122.3 kB, resulting from the 61,148 parameters used in 16-bit precision (fp16). 

The baseline's accuracy on the development-test split ranges from 42.40\% for the smallest training subset (5\%) to 56.99\% accuracy for the full set (100\%). The performance increases monotonically as the number of audio segments available for training increases. The code and a detailed description of the baseline system are available online\footnote{Source Code: \url{https://github.com/CPJKU/dcase2024_task1_baseline/tree/main}}.


\vspace{-4pt}
\section{Challenge Results}
\label{sec:results}
\vspace{-4pt}

The task received 37 submissions from 17 teams and is therefore the second most popular task in the 2024 edition of the DCASE challenge. The slight decrease in popularity compared to the previous year's edition is likely due to the more complex setup. 16 out of 17 teams outperformed the baseline system and for most of the teams, the performance on the development-test split aligns well with the performance on the evaluation set. The challenge website contains detailed results and descriptions on all submitted systems\footnote{Results: \url{https://dcase.community/challenge2024/task-data-efficient-low-complexity-acoustic-scene-classification-results}}.

Table~\ref{tab:results} presents the best systems submitted by the ten top-ranked teams and lists details in terms of architectures, complexity handling, device generalization, and usage of external resources. \textit{Score} denotes the average accuracy across all five training set splits on the evaluation set. Note that a team's rank depends on all three allowed submissions, rather than only on the system achieving the highest score (which is why the \textit{Team Rank} column of Table~\ref{tab:results} is not perfectly sorted).


\vspace{-3pt}
\subsection{Architectures}
\vspace{-3pt}

In response to the low-complexity constraints and following the trend observed in the previous edition of this task~\cite{Martin22asc_in_dcase22}, the large majority of systems are based on factorized CNN architectures. Most prominently, factorization is realized via inverted residual blocks, as used in the baseline architecture. Table~\ref{tab:results} shows that four out of the ten best systems are based on modified versions of the CP-Mobile architecture~\cite{Schmid2023cpm}. The top-ranked system~\cite{Han2024rank1} further reduces CP-Mobile's complexity by factorizing the spatial convolutions with a 3x3 kernel into two separate convolutions with 1x3 and 3x1 kernels. Team \textit{Shao\_NEPUMSE}~\cite{Shao2024rank2} enhances an inverted residual block-based architecture with parallel Mamba blocks~\cite{Gu23mamba}, a derivative of state space models. Teams \textit{Cai\_XJTLU}~\cite{Cai2024rank5} and \textit{Park\_KT}~\cite{Park2024rank6} use modified versions TF-SepNet~\cite{Cai23tfsepnet} and BCRes2Net~\cite{Kim2021qti}, respectively, both of which achieved high ranks in previous editions of this task and decouple spatial convolutions over frequency and time dimensions. Team \textit{OO\_NTUPRDCSG}~\cite{Oo2024rank7} introduces MofleNet by enhancing the CP-Mobile architecture with channel shuffle operations; Team \textit{Truchan\_LUH}~\cite{Truchan2024rank9} uses an isotropic convolutional architecture following a patch embedding layer; and Team \textit{Yan\_NPU}~\cite{Yan2024rank10} presents MAR-CNN, an asymmetric multi-branch convolutional architecture. 


\vspace{-3pt}
\subsection{System Complexity}
\vspace{-3pt}

Knowledge Distillation (KD) can be identified as the most prominent technique to tackle the low-complexity constraints, with the six top-ranked teams using KD. The most popular teacher model is the audio spectrogram transformer PaSST~\cite{koutini22passt}. Among other models that proved to be successful teachers are CP-ResNet~\cite{koutini21cpr} (Teams \textit{MALACH24\_JKU}~\cite{David2024rank2} and \textit{Shao\_NEPUMSE}~\cite{Shao2024rank2}), BEATs~\cite{chen23beats} (Team \textit{Cai\_XJTLU}~\cite{Cai2024rank5}), EAT~\cite{Chen24EAT} (Team \textit{Park\_KT}~\cite{Park2024rank6}) and DyMN~\cite{Schmid24DyMN} (Team \textit{Bai\_JLESS}~\cite{Bai2024rank14}). 
Regarding numerical representation of parameters, both 8-bit and 16-bit precision solutions are among the top-ranked systems. To convert parameters to 8-bit precision, Teams \textit{Shao\_NEPUMSE}~\cite{Shao2024rank2} and \textit{OO\_NTUPRDCSG}~\cite{Oo2024rank7} use Quantization-Aware-Training, while Team \textit{Cai\_XJTLU}~\cite{Cai2024rank5} shows that also Post-Training Static Quantization can lead to good results. 
In addition to KD, the top-ranked system by Team \textit{Han\_SJTUTHU}~\cite{Han2024rank1} uses pruning. They construct a large version of SSCP-Mobile by increasing the number of channels, apply pruning to meet complexity constraints, and then fine-tune the pruned model using KD. 

\vspace{-3pt}
\subsection{Device Generalization}
\vspace{-3pt}

The majority of teams tackle the device mismatch with dedicated data augmentation techniques. In this regard, Freq-MixStyle~\cite{Kim2022FMS,Schmid2022CPJKU22}, which is also integrated into the baseline system, is used by all of the ten top-ranked teams. Additionally, seven out of the ten top-ranked systems use device impulse response augmentation, implemented using convolution with 66 freely available impulse responses from MicIRP\footnote{\url{http://micirp.blogspot.com/}}. 
An interesting alternative is presented by Team \textit{Truchan\_LUH}~\cite{Truchan2024rank9} using an adversarial device classifier that forces the feature extractor to learn device-invariant representations. 

\begin{figure}[t]
\centering
{\includegraphics[width=\linewidth]{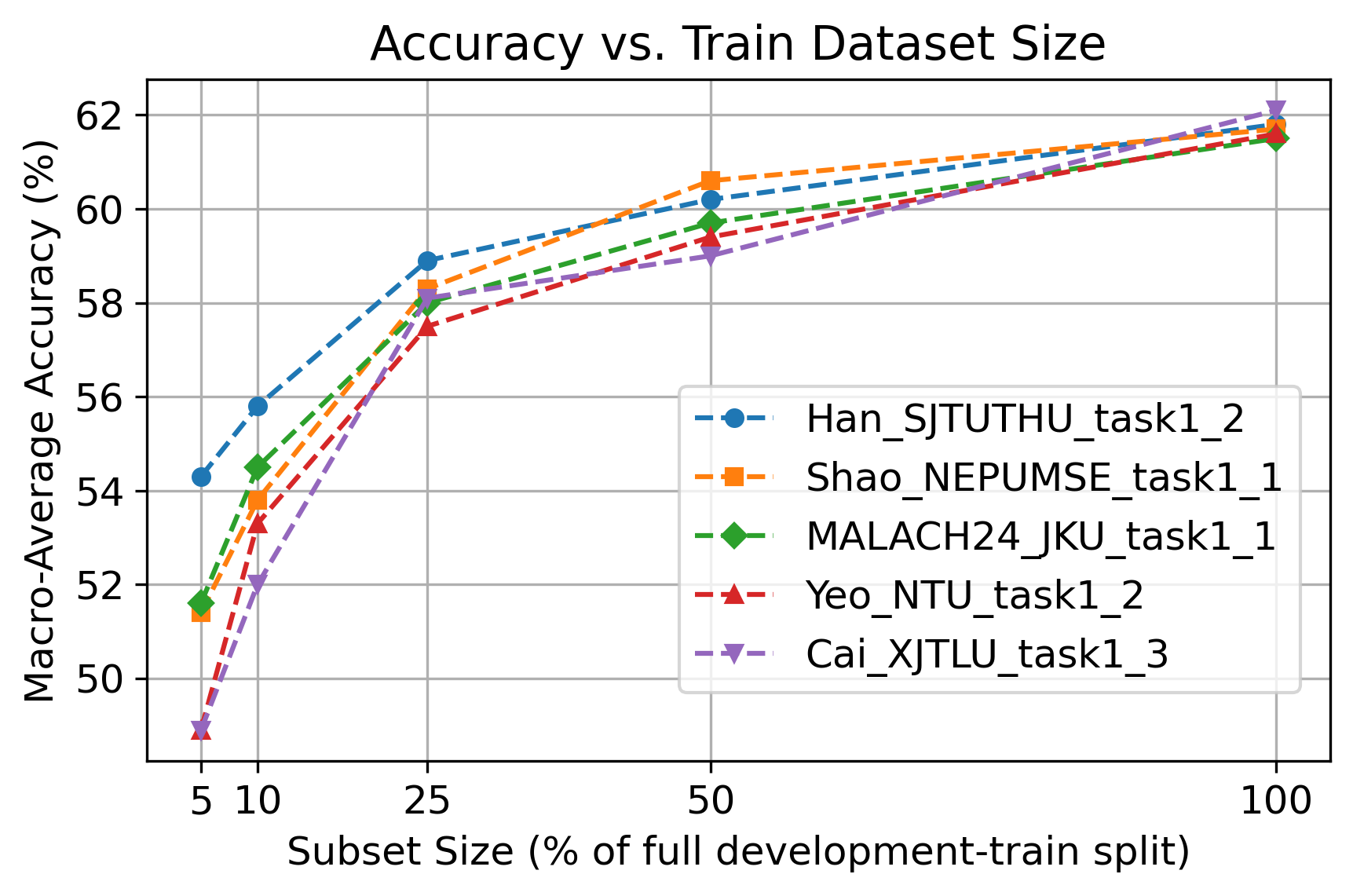}}
\caption{Performance of the best systems from the five top-ranked teams on the evaluation set for training on the five subsets (5\%, 10\%, 25\%, 50\%, 100\%) of the development-train split.}
\vspace{-12pt}
\label{fig:top5_systems}
\end{figure}

\vspace{-3pt}
\subsection{Limited Training Data}
\vspace{-3pt}

Figure~\ref{fig:top5_systems} shows that the top systems submitted to the challenge perform similarly when trained on the 100\% train split. However, the smaller the size of the training set, the larger the performance differences, underscoring the large impact of creating data-efficient systems. In fact, the top-ranked system does not achieve the highest accuracy for the 50\% and 100\% training splits, but it surpasses other systems on the 5\%, 10\%, and 25\% subsets. In particular, on the smallest training set, it outperforms all other teams' systems by more than 2 percentage points in terms of accuracy.

In the following, we describe approaches by participants to counteract the performance dropoff for small training sets.

\textbf{General-Purpose Audio Datasets:} Very commonly, participants make use of large general-purpose audio datasets, in particular, AudioSet~\cite{Gemmeke17audioset}, to alleviate the data scarcity problem. This is achieved in three different ways: (1) by fine-tuning a large, pre-trained model on ASC and using it as a teacher model in a KD setup; (2) by directly pre-training a low-complexity model on AudioSet; and (3) by extracting audio clips from AudioSet as additional training data. 
The effectiveness of (1) is underlined by the fact that most of the top-ranked teams use an AudioSet pre-trained transformer model as a teacher in a KD setup. For example, Team \textit{Cai\_XJTLU}~\cite{Cai2024rank5} achieves an accuracy of 55.7\% on the development-test set when fine-tuning multiple BEATs~\cite{chen23beats} models on the 5\% training subset, which is higher than the Baseline system's accuracy using 10 times as much training data. Regarding (2), the team with the second-best performance on the 5\% and 10\% subsets, Team \textit{MALACH24\_JKU}~\cite{David2024rank2}, pre-trains CP-Mobile on AudioSet and reports a large performance gain for fine-tuning on smaller training subsets. Concerning (3), Team \textit{Werning\_UPBNT}~\cite{Werning2024rank8} trains a dataset domain classifier to extract audio clips from AudioSet that are similar to the samples in the respective training sets and uses these as additional samples for KD. Additionally, Team \textit{Surkov\_IMTO}~\cite{Surkov2024rank13} selects AudioSet clips from specific event classes such as \textit{Bus} or \textit{Train} and uses them as additional unlabeled samples in a mean-teacher approach.

\textbf{Extensive Data Augmentation:} Besides Freq-MixStyle and DIR augmentation, extensive data augmentation is applied to improve generalization performance on the small training sets. In this regard, Team \textit{MALACH24\_JKU}~\cite{David2024rank2} uses FilterAugment~\cite{Nam22FA}, Team \textit{Shao\_NEPUMSE}~\cite{Shao2024rank2}) experiments with audio playback, Team \textit{Chen\_SCUT}~\cite{Chen2024rank12} uses Spectrum Modulation. SpecAugment, time rolling, and Mixup are widely used throughout submissions.

\textbf{Model Size and Architecture:} Team \textit{Yeo\_NTU}~\cite{Yeo2024rank5} investigated the relationship between model size and performance on small training splits and found that models of reduced complexity generalize better for small training splits. Team \textit{Park\_KT}~\cite{Park2024rank6} enhanced their network with Snake activation functions and showed that the introduced inductive bias on periodicity leads to a large performance gain on smaller training sets.

\vspace{-4pt}
\section{Conclusion}
\label{sec:conclusion}
\vspace{-4pt}

This paper has presented an analysis of Task 1 in the DCASE 2024 challenge, which focused on the real-world deployment of ASC systems with low-complexity constraints, device mismatch, and training data scarcity being the main hurdles to overcome. The task remained the second most popular in the DCASE 2024 challenge, underscoring the high interest in the task despite the increasingly challenging setup. Multiple strategies have been proposed to tackle the limited availability of training data; most highly-performing systems transferred knowledge from a large general-purpose audio dataset to the ASC task, either in the form of pre-trained models or by extracting additional ASC-related audio clips for training. Data augmentation remained a highly important aspect, not only to address device generalization but also to improve generalization capabilities, with only a small training set available. Other solutions to the data scarcity problem involve adapting the model's complexity or building inductive biases into the model architecture. Summarizing the output of the task, several promising techniques have been proposed that can boost performance on downstream tasks when only a small training set is available. 

\vspace{-4pt}
\section{ACKNOWLEDGMENT}
\label{sec:ack}
\vspace{-4pt}

The LIT AI Lab is supported by the Federal State of Upper Austria. Gerhard Widmer's work is supported by the European Research Council (ERC) under the European Union's Horizon 2020 research and innovation programme, grant agreement No 101019375 (Whither Music?).


\bibliographystyle{IEEEtran}
\bibliography{refs}

%
%
%
%
%
%
%
%
%

\end{sloppy}

\end{document}